**Testing metabolic scaling theory using intraspecific allometries in Antarctic microarthropods.**

Tancredi Caruso[*, 1], Diego Garlaschelli[2], Roberto Bargagli[1], Peter Convey[3]

[1] Department of Environmental Sciences "G. Sarfatti", University of Siena, via P.A. Mattioli, 4, 53100 Siena, Italy.

[2] Department of Physics, University of Siena, via Roma, 56, 53100 Siena, Italy.

[3] British Antarctic Survey, Natural Environment Research Council, High Cross, Madingley Road, Cambridge CB3 0ET, UK

* Corresponding author: phone +39 0577232877, fax +39 0577232806, tancredicaruso@unisi.it

## ABSTRACT

Quantitative scaling relationships among body mass, temperature and metabolic rate of organisms are still controversial, while resolution may be further complicated through the use of different and possibly inappropriate approaches to statistical analysis. We propose the application of a modelling strategy based on the theoretical approach of Akaike's information criteria and non-linear model fitting (nlm). Accordingly, we collated and modelled available data at intraspecific level on the individual standard metabolic rate of Antarctic microarthropods as a function of body mass (M), temperature (T), species identity (S) and high rank taxa to which species belong (G) and tested predictions from Metabolic Scaling Theory (mass-metabolism allometric exponent $b$ = 0.75, activation energy range 0.2 - 1.2 eV). We also performed allometric analysis based on logarithmic transformations (lm). Conclusions from lm and nlm approaches were different. Best-supported models from lm incorporated T, M and S. The estimates of the allometric scaling exponent linking body mass and metabolic rate indicated no interspecific difference and resulted in a value of 0.696 ± 0.105 (mean ± 95% CI). In contrast, the four best-supported nlm models suggested that both the scaling exponent and activation energy significantly vary across the high rank taxa (Collembola, Cryptostigmata, Mesostigmata and Prostigmata) to which species belong, with mean values of $b$ ranging from about 0.6 to 0.8. We therefore reached two conclusions: 1, published analyses of arthropod metabolism based on logarithmic data may be biased by data transformation; 2, non-linear models applied to Antarctic microarthropod metabolic rate suggest that intraspecific scaling of standard metabolic rate in Antarctic microarthropods is highly variable and can be characterised by scaling exponents that greatly vary within taxa, which may have biased previous interspecific comparisons that neglected intraspecific variability.

## INTRODUCTION

Unifying concepts such as scaling and fractal geometry have been receiving increasing attention in the biological and ecological literature (Garlaschelli et al. 2003; Brown et al. 2004, 2005; Kozlowski and Konarzewski 2004; Chown et al. 2007; Makarieva et al. 2008), as they suggest a consistent picture encompassing various levels of description, and at the same time provide simple relationships for key functional quantities. In particular, metabolic scaling theory (MST, Brown et al. 2004, 2005) provides simple relationships linking geometrical properties (the volume or mass of an organism) to biological processes (metabolic rate) and environmental conditions (temperature). Indeed, biological allometry has received new interest from theoretical models such as MST that elegantly explain the 3/4 power scaling of metabolic rate with body mass (West et al. 1997). Since Huxley (1932), power functions of the form $Y = Y_0 M^b$ have been proposed to describe a biological rate (Y) that has some power function of organism body mass (M). The allometric scaling exponent *b* ruling such power functions has been claimed to have the unusual property of being a multiple of 1/4 rather than the most conventional Euclidean 1/3, which for example links length to volume. Amongst the best known of allometric relationships is that of Kleiber (1932), in which individual metabolic rate *I* scales as $I = I_0 M^{3/4}$. MST explains the allometric exponent and more generally, the predominance of quarter law allometric scaling, by assuming geometrical limitation in the rates of uptake of resources and the distribution of materials through organism branching networks, which would behave as fractal objects (West et al. 1997, 1999; Savage et al. 2004). Models like MST are said to be universal because they are based on physical first principles, which are applicable both within and across species and lead to scaling parameters that do not vary across individual, species and higher rank taxa (Glazier 2005; Price et al. 2009). Since the first formulation of MST, models describing metabolic rates have been extended to include temperature effects, as described by the Boltzmann factor $e^{-E/kT}$, where *k* is Boltzmann's constant (8.62 x $10^{-5}$ eV $K^{-1}$), *E* is the activation energy of the biochemical reaction and *T* is absolute temperature (Gillooly et al. 2001; Brown et al. 2004). Following Marquet et al. (2004), stoichiometry (*R*), interpreted as an integration of resources and their relative proportions, has been introduced in order to obtain the general metabolic model, $I = I_0 M^{3/4} e^{-E/kT} f(R)$, which reasonably assumes that stoichiometric effects have multiplicative effects on metabolism, although the function *f* has not been defined (Sterner and Elser 2002; Gillooly et al. 2002).

Although MST was proposed as a unifying theory underlying the structure and function of ecosystems, several authors have raised important theoretical and empirical concerns regarding its validity (e.g. Glazier 2005; Makarieva et al. 2005, 2008; Chown et al. 2007). One of the most interesting elements of this debate focuses on the validity of the empirical patterns the theory attempts to explain (Makarieva et al. 2008; Packard and Birchard 2008; Packard 2009). The methodological caveat recently pointed out by Packard and colleagues (e.g. Packard and Birchard 2008; Packard 2009) is particularly interesting, highlighting that most investigators have not validated allometric equations in the original scale of measurements. Instead, allometric equations have been linearised by logarithmic transformation, and linear models then employed for the estimation of relevant allometric scaling exponents (e.g. Kleiber 1932; Brown et al. 2004; Savage et al. 2004; Glazier 2005). As they have shown (e.g. Packard 2009) through synthesising information available from methodological literature (e.g. Osborne and Overbay 2004; Warton et al. 2006), there are at least four sources of bias in this 'traditional' approach: 1, the non-linear logarithmic transformation alters the relationships between dependent and independent variables, in particular reducing the influence of outliers; 2, if an equation with a non-zero intercept provides a better fit to the original data, parameters estimated by linear regression after logarithmic transformation are misleading; 3, the underlying statistical model assumes multiplicative errors, which usually is not appropriate when equations are back-transformed to the original arithmetic scale; 4, when using most common procedures, such as ordinary least squares, small values have much greater influence than large values on parameter estimates. These points may have dramatic consequences on parameter estimates. For instance, Packard and Birchard (2008) reanalysed data for the basal metabolic rate of 626 species of mammals (sourced from Savage et al. 2004) and found that a straight line fitted to log-transformed data does not satisfy the assumptions underlying the analysis, and that the allometric equation then obtained by back transformation underestimates BMR for the largest species because the estimation of scaling exponents was strongly biased by the log-transformation. However, Kerkhoff and Enquist (2008) countered this, asserting that multiplicative errors by logarithmic transformation are instead appropriate for studying biological allometry since this is usually assumed to be originated by multiplicative processes (e.g. growth). Therefore, the debate remains to be resolved, while there have been a very few attempts (e.g. Packard and Birchard 2008) to compare results obtained from linear models applied to logarithmic (transformed) data and nonlinear models applied to data in the original scale of measurements.

Several recent broad-scale meta-analyses of metabolic characteristics of non-polar invertebrates (Meehan 2006) and higher insects (Addo-Bediako et al. 2002; Chown et al. 2007; Makarieva et al. 2008) have explored fundamental patterns of scaling in metabolism, biomass and temperature. However, the application of such approaches to Antarctic soil microarthropods or invertebrates generally is still lacking and, in particular, no studies have applied a modelling approach based on non-linear regressions and model selection criteria such as Akaike's criterion (Burnham and Anderson 2002; Johnson and Omland 2004, Duncan et al. 2008) to attempt to unravel methodological and theoretical issues that are typically inherent in the patterns obtained in analyses of ecophysiological data. Also, many studies of arthropods do not separate intra- and interspecific scaling of metabolic rate, mainly because the former is usually underestimated in the allometric analysis of metabolism (Brown et al. 1997; Glazier 2005). Theoretically, if the assumption of low intraspecific variability is valid, interspecific analysis can ignore intraspecific variability when testing the applicability of universal models such MST. Indeed, universal models predicting the same scaling exponents are valid across individuals, populations, species and phyla. However, evidence suggests that some published interspecific comparisons may be biased by high variability at intraspecific levels, and Glazier (2005) strongly recommended that future research should give greater recognition to intraspecific variability. In principle, intraspecific and interspecific allometries differ and, while testing for intraspecific variability can provide a valid test of universal models, results from intraspecific analysis do not directly apply to interspecific relationships (Price et al. 2009). Therefore, data that provide evidence of high intraspecific variability challenge experimentally some of the assumptions of universal models such as MST, even though they do not themselves address metabolic allometries at interspecific levels (Glazier 2005; Price et al 2009).

Physiological studies relevant to these questions are available as part of the considerable literature addressing ecophysiological stress tolerance strategies that has been a focus of Antarctic biological research over at least the last three decades. Physiological ecology has a key role in the understanding of macroecological patterns and ecosystem functioning (Blackburn and Gaston 2003). This statement has particular strength for Antarctic terrestrial ecosystems, characterised by low biodiversity, simple food web structures, the overwhelming dominance of a few invertebrate groups (especially arthropods), and the predominance of physical over biological variables as environmental selective pressures (Convey 1996; Chown and Gaston 1999; Peck et al. 2006). Because organism metabolism is the fundamental crossroad of energy fluxes through ecosystems (Brown et al. 2004; Makarieva et al. 2005), it is very important to determine the relationships that link metabolic rate to fundamental biological characteristics such as biomass and environmental parameters such as temperature. Furthermore, understanding of these linkages is a fundamental component of being able to predict likely responses to current trends and scenarios of global and regional environmental change (Chown and Convey 2006; Wall 2007). Here, again, understanding of polar and particularly Antarctic ecosystems is important in this context, as parts of the continent are experiencing rates of change that are the fastest currently seen on the planet (Turner et al. 2005; Convey 2006).

Within this context, the present study compiled existing data in order to model metabolism of Antarctic microarthropods as a function of body mass, temperature and taxonomic identity. As with analogous studies (e.g. Meehan 2006), models were constructed under the framework of metabolic scaling theory, the predictions of which were then evaluated and compared to predictions from other models (e.g. Peters 1983; Glazier 2005; Chown et al. 2007; Makarieva et al. 2008, Price et al. 2009), including the classic euclidean scaling (mass-metabolism allometric exponent $b$ = 0.66) and the cell size model ($0.67 \leq b \leq 1$). The dataset analysed includes multiple data for various Antarctic species, allowing the analysis to be performed at the intraspecific level.

## METHODS

### Data collection

A database was constructed based on extensive bibliographic research performed via both web-based and library sources (e.g. see Block 1992). This initially identified ~ 100 publications as being potentially relevant based on the following keywords: metabolism, biomass, mass, temperature, rate, Antarctic, polar, arthropods, microarthropods, soil, terrestrial, oxygen, consumption, cold, adaptation. After a full examination of this literature, a more limited set of published studies (Block and Tilbrook 1975, 1978; Block 1976, 1977, 1979; Goddard 1977a,b; Procter 1977; Young 1979) were identified as being relevant to and fulfilling the requirements of the planned analyses. Data on metabolism in these studies were collected by a single method, Cartesian diver respirometry, removing the potential confounding effect of use of different methodologies. These studies primarily describe research performed at Signy Island (South Orkney Islands, maritime Antarctic), and to a lesser extent sub-Antarctic South Georgia. While several other studies reported clear information on metabolism and temperature they did not consider biomass. In these published studies, metabolic rates were expressed as oxygen consumption rates, usually as µL $g^{-}$

$^{1}$ $h^{-1}$. These were converted into mL $ind^{-1}$ $h^{-1}$ on the basis of information and equations provided in the original studies, and then oxygen consumption rate was converted into metabolic rate expressed as J $h^{-1}$ through the oxyenergetic coefficient of 20.20 J $mL^{-1}$ $O_2$ (Prus 1975; Meehan 2006). Following the procedure adopted by Meehan (2006), measures of mean metabolic rates and body mass were entered separately for each life stage, sex and temperature level.

**Modelling**

Metabolic scaling theory provides a synthetic framework that allows the modelling of individual metabolism (*I*) as a function of both body mass (*M*) and absolute temperature (*T*) according to the equation $I = I_0 M^b e^{-E/kT}$, where the allometric exponent *b* is predicted to be 0.75, *k* is Boltzmann's constant (8.62 x $10^{-5}$ eV $K^{-1}$) and *E* is the activation energy of the biochemical reaction, which is expected to be between 0.2 and 1.2 eV (Gillooly et al. 2001). The equation is simply linearised by log transformation. Linear models (lm) have thus been traditionally employed, with parameters then estimated by least squares regression. Here, following Packard and Birchard (2008), we performed both an analysis based on logarithmically transformed data (lm) and a non-linear analysis using non-transformed data (nlm).

Competing models were defined as follows:

1) generic allometric model with biomass parameter only (Kleiber's law): $I = I_0 M^b$
2) as (1) with a species (S) or taxonomic group (G) additive effect on the prefactor $I_0$: $I = I_0[S]M^b$ and $I = I_0[G]M^b$ (i.e. the prefactor $I_0$ differs between species [S] or high rank taxa [G]).

3) general metabolic scaling theory model with parameters for biomass and temperature (Gillooly et al. 2001; Brown et al. 2004): $I = I_0 M^b e^{-E/kT}$

4) as (3) with a species or taxonomic group conditional linear effect on the prefactor $I_0$: $I = I_0\left[S\right] M^b e^{-E/kT}$ and $I = I_0\left[G\right] M^b e^{-E/kT}$

5) as (3) but with the allometric exponent varying among species or taxonomic group:

   $I = I_0 M^{b[S]} e^{-E/kT}$ and $I = I_0 M^{b[G]} e^{-E/kT}$

6) as (5) but with the activation energy varying among species or taxonomic group:

   $I = I_0 M^b e^{-E[S]/kT}$ and $I = I_0 M^b e^{-E[G]/kT}$

7) as (5) but with both the allometric exponent and activation energy varying among species or taxonomic group:

   $I = I_0 M^{b[S]} e^{-E[S]/kT}$ and $I = I_0 M^{b[G]} e^{-E[G]/kT}$

Higher taxonomic groups considered (G) were Collembola, Cryptostigmata, Prostigmata and Mesostigmata. At this taxonomic level, different strategy usage may exist in response to the extreme conditions of Antarctic ecosystems, and such differences may be responsible for significant differences in metabolism among major microarthropod groups (e.g. Convey 1996, 1997; Peck et al. 2006). In the past, allometric analysis of metabolism focused on interspecific comparison and assumed that intraspecific variability in scaling exponent can be disregarded because in most intraspecific analysis 3/4 is within the statistical confidence interval of scaling exponent estimates (Brown et al. 1997). However, Glazier (2005) recently reported compelling evidence that this assumption is far from being confirmed. Our models 5 and 7 allow the scaling exponent to vary among taxa. This procedure is analogous to that employed by Price et al. (2009), who also tested for models that do not make specific predictions for scaling exponent and could also result in species-specific scaling parameters. In an analogous fashion, models 6 and 7 test for activation energy varying among taxa. Model 7 additionally allows both the scaling exponent and activation energy to vary.

Theoretically, variations in the prefactor (Brown's normalization constant) may depend on several factors (Brown et al. 2004), including the multiplicative effects of factors not explicitly accounted for by the general MST model with Mass and Temperature only. For instance, high variability in $I_0$ could reflect the role played by stoichiometry (e.g. Sterner and Elser, 2002) $R$ because $I = I_0 M^{3/4} e^{-E/kT} f(R) \rightarrow I = sM^{3/4} e^{-E/kT}$ with $s = I_0 E(f(R))$ and $E(f(R))$ being some average expectation of $f(R)$. If species or taxa are significantly different in their expectation for $f(R)$, but the latter function is not explicitly accounted for in models, the multiplicative factor $s$ may include both variation in $I_0$ and stoichiometry. It is thus interesting to explore at which taxonomic level the multiplicative factor may vary. For instance, strong phylogenetic signals are usually observed in parameters of allometric relationships, including those linking metabolism and mass. This means that methodology able to account for residual correlation due to species phylogenetic proximity is logically required (Duncan et al. 2004). In the context of the current study, we were not able to account for this effect given the lack of relevant molecular data. However, given the taxonomic composition of the analysed dataset (see Results, below) it is also reasonable to assume that phylogenetic auto-correlation is likely to have a minor influence here. Another problem is that some of higher taxonomic groups considered in this study only included single species, which could in part bias the analysis. However, in order to gain preliminary and necessary information on this point, we nevertheless considered these models given the low diversity of the analysed system and the fact that this is the only dataset available for polar regions. The conclusions from taxonomic aggregation were then considered in the light of this limit of the dataset.

The procedure nls of R (version 2.8.1: http://www.r-project.org.) was used for determining the non-linear least-squares estimates of the parameters fitting non-linear models (nlm; R Development Core Team 2006; Ritz and Streibig 2008). Models were also linearised by log-transformation and fitted to logarithmically transformed data (lm) by the procedure lm of R.

Within the two sets of analysis (nlm and lm) model assessment and selection relied on a theoretical approach using Akaike's criterion corrected for sample size ($AIC_c$: Burnham and Anderson 2002; Johnson and Omland 2004). Models were accordingly ranked, and the model with the minimum $AIC_c$ (below $AIC_{min}$) was used as the reference for calculating AIC difference ($\Delta_i$) and model weights ($w_i$). Models within 2 AIC units from $AIC_{min}$ were considered competitive and most plausible, and their model weights provided an estimate of their strength (Burnham and Anderson 2002).

## RESULTS

The compiled database included 130 metabolic rates that synthesise data obtained from several hundred measurements of oxygen consumption rate across 9 species (Appendix S1): two springtails (Collembola), one oribatid mite (Acari: Cryptostigmata), five prostigmatid mites (Acari: Prostigmata) and one predatory mite (Acari: Mesostigmata). The dataset is taxonomically and geographically limited but, considering the very low biodiversity typical of Antarctic terrestrial ecosystems, it is representative of arthropod taxonomic and functional diversity in the region. Body mass ranged from 0.04 μg in smaller species and younger life stages to 214 μg in adults of larger species, while metabolic rates ranged from 6.6E-07 J $h^{-1}$ to 0.002 J $h^{-1}$.

Among nlms, ranking based on Akaike's criterion (Table 1) selected

$I = I_0 M^{b[G]} e^{-E/kT}$, $I = I_0 M^{b} e^{-E[G]/kT}$, $I = I_0 \left[G\right] M^{b} e^{-E/kT}$

as the strongly superior models with a sum of weights of about > 0.90 and Δ $AIC_c$ within two units from the best model ($I = I_0 M^{b[G]} e^{-E/kT}$). Then the model $I = I_0 M^{b[G]} e^{-E[G]/kT}$ followed, with Δ $AIC_c$ from the best model = 2.14 and Akaike's weight = 0.104. Δ $AIC_c$ of all other models was > 10. The sum of the Akaike's weight of the above four models = 0.997 and, therefore, there was very weak support for all other competing models (Burnham and Anderson 2002). Thus, according to the four best models, each major taxonomic group (G) requires a specific prefactor ($I_0$), or scaling exponent (*b*), or activation energy (*E*), or both scaling exponent and activation energy, although the latter had a slightly weaker support. The worst-fitting model was the Kleiber's law.

Considering log-transformed data (Table 2), the best-supported lm was

$\log(I) = \log(I_0) + S + b\log(M) + (E / kT)$

with a weight > 0.99. The Δ $AIC_c$ from the second best model being > 11 indicates that all other competing models offered a very poor fit relative to the most informative (Burnham and Anderson 2002). In contrast with the nlm, the best lm indicated that each species requires a specific prefactor while species did not significantly vary for the scaling exponent and activation energy. Parameter estimates and standard errors of the best-fitting model indicate that 95 % CI for the scaling exponent of body mass (*b*) and activation energy *E* were 0.591 to 0.801 and 0.302 to 0.673, respectively. The worst-fitting model was again Kleiber's law.

Generally, for models assuming that allometric scaling exponent *b* and activation energy do not vary among species, the errors associated with the estimate of *b* were larger for non-linear than linear models (Table 3).

Considering the best ($\Delta$ $AIC_c = 0$) nlm, the estimates of *b* were very variable among the four major taxa (Table 4), this variability being inconsistent with both the expectation from MST and Euclidean scaling (Fig. 1) of a universal scaling exponent.

**DISCUSSION**

Our model results do not refute the metabolic scaling theory prediction of a scaling exponent *b* = 0.75, and for the most plausible models identified by both nlm and lm the 95% CI included 0.75. However, the estimate obtained was also compatible with the classical Euclidean expectation of *b* = 0.66 (White and Seymour 2003) and models that predict a highly variable *b*, in particular the cell-size model that predicts *b* ranging from 0.67 to 1 for lower taxonomic groups and within species (Chown et al. 2007). Indeed, the error associated with the estimates of the scaling exponent here is large compared to that reported in several other datasets (Addo-Bediako et al., 2002; Savage et al. 2004; Meehan 2006; Packard and Birchard 2008), and our data and modelling clearly indicate that this variability was mainly underlain by the highly variable scaling exponent observed within taxa (e.g. Fig.1 and Table 4).

The predictions of metabolic scaling theory in relation to activation energy (*E*) include a large range (0.2 - 1.2 eV; Gillooly et al. 2001), and our observed 95% CI is well centred within this range, in respect of both the nlm and lm approaches.

However, nlm again showed that the data were consistent with the hypothesis that mean activation energy varies among major taxonomic groups (e.g. see Table 2, row 2, the model $I = I_0 M^b e^{-E[G]/kT}$). However, this is a preliminary conclusion that could potentially be driven/biased by Collembola and Prostigmata, the only two higher taxa represented by more than one species. Nevertheless, if the species effect was very large within Prostigmata (five species) and Collembola (two species), the likelihoods of models such as $I = I_0 M^b e^{-E[S]/kT}$ or $I = I_0 M^{\beta[\Sigma]} \varepsilon^{-E[\Sigma]/\kappa T}$ would be greater and able to compensate for the relatively larger number of parameters (more parameters for *S* than *G*) that caused their penalisation in the AIC.

In line with Glazier (2005) and the recent modelling approach of Price et al. (2009), the allometric analysis of our collated dataset strongly indicates that the assumption of low intraspecific variability made by some interspecific analysis (e.g. Addo-Bediako et al. 2002; Meehan 2006) does not always apply to arthropods.

Our analyses confirm the importance of two often neglected facts: first, log transformation may itself alter the final conclusion drawn from statistical analysis (Packard and Birchard 2008; Packard 2009); second, classical physiological studies rapidly recognised large inter- and intra-taxonomic variability (e.g. Precht et al. 1979; Glazier 2005), that has often been neglected by universal models in the attempt to search for general and broad scale pattern in the allometric relationships between metabolic rate, biomass and temperature. Further, we also suggest that this is due to the adopted modelling strategy, itself the subject of continued methodological debate within the relevant scientific communities (e.g. Addo-Bediako et al. 2002; Hodkinson 2003; White and Seymour 2003; Brown et al. 2004; Kozlowski and Konarzewski 2004; Brown et al. 2005; Makarieva et al. 2005, 2008; Price et al. 2009).

For instance, Makarieva et al. (2005, 2008), in an interspecific approach, challenged on statistical and methodological grounds the universality of the allometric scaling exponent of 0.75, suggesting that the calculated scaling exponent may instead arise from errors in data assortment and analysis. When data are sorted based on biologically reasonable criteria, such as separating unicellular organisms, invertebrates and endotherms, and the most reasonable temperature corrections are performed, the expected global -0.25 scaling exponent (which should regulate mass-specific metabolic rate $q$) is questionable (Makarieva et al. 2005). Separating different taxa is a way to account for intrataxonomic differences when comparing different taxa. Indeed, Makarieva et al. (2008) provide an analysis suggesting that mean mass-specific metabolic rates have only a thirty-fold variation across life's disparate forms, while the generally-used allometric scaling laws would predict a range from 4,000 to 65,000-fold.

In summary, data show that each group of Antarctic arthropods has its own specific constants that need to be experimentally estimated prior to extrapolation of metabolic rates based on body mass data, environmental temperature and other traits such as population density. The relatively large errors and taxonomic variability associated with the estimates of scaling parameters and activation energy indicate the need for more precise measurement and suggest that the assumption of universal models may not apply to arthropods (Price et al. 2009). Further intraspecific variations not accounted for by our models could take the form of age or life stage effects. Indeed, existing Antarctic data highlight a further important, but often neglected, finding of both the earlier literature and more recent reviews (*e.g.* Block and Tilbrook 1975; Block and Young 1978; Young 1979; Precht et al. 1979; Hodkinson 2003; Glazier 2005; Chown et al. 2007), that emphasizes the large metabolic variations observed between different life stages of a single species that are characterised by differential energetic investment in growth and maintenance costs.

The existence of significant differences in metabolism among major microarthropod groups (at least for Collembola and Prostigmata, the only two taxa here represented by at least two species) implies links with different strategy usage in response to the extreme conditions of Antarctic terrestrial ecosystems (cf. Convey 1996, 1997; Peck et al. 2006). For instance, differential investment in elements of life history and ecophysiological strategy will imply differential energetic investment in biochemical pathways such as production of antifreeze compounds or, more generally, in cryoprotectant mechanisms (Peck et al. 2006), with consequential effects on other elements of life history strategy (e.g. Convey 1996; Convey et al. 2003; Hennion et al. 2006). In this respect, stoichiometry (R) could play a key role as suggested by the general metabolic equation $I = I_0 M^{3/4} e^{-E/kT} f(R)$ (Marquet et al. 2004), and we hypothesise that large taxonomic variations in the prefactor may actually reflect variation in the function accounting for *R*. Stoichiometry (Sterner and Elser 2002) is a general concept that integrates the various ecological concepts relating organismal requirements for resources, which generally differ between taxa and their macro- and micro-environments, especially in Antarctica (Convey 1996; Chown and Gaston 1999; Peck et al. 2006).

Elevation of metabolic rate at low temperature (metabolic cold adaptation, MCA; Cannon and Block 1988; Block 1990; Convey 1996; Peck et al. 2006) is proposed as a frequent but not ubiquitous element of the evolutionary response to the low thermal energy availability typical of Antarctic terrestrial ecosystems. It appears to include reduction in enzyme activation energies and a disproportionate response of reaction rate to small temperature increments at low temperature ($Q_{10}$) (Block 1990; Peck et al.

2006), both argued to allow rapid advantage to be taken of any slight increase in temperature for resource exploitation. However, new data on the continental Antarctic springtail *Gomphiocephalus hodgsoni* (McGaughran et al. 2009) provide a novel suggestion of clear intra-seasonal and temperature-independent variation in mass-specific standard metabolic rate. This implies that metabolic rate can also be related to the seasonal tuning of biological activity rather than simply being a response to temperature *per se*, an interpretation that is well documented in the thermally stable Antarctic marine environment (see Peck et al. 2006), but that adds a further level of complexity to attempts to model allometric functions for these taxa.

We conclude that MST remains a good starting framework for modelling the mass and temperature dependence of metabolism in animals, including Antarctic microarthropods, but that much theoretical (e.g. Price et al. 2009) and experimental work is needed in order to be able to generate adequate indirect estimates of metabolic rate, and permit its application to the study of energy and mass fluxes through the terrestrial ecosystem.

## ACKNOWLEDGEMENTS

This study is a contribution to the EBESA IPY project n° 452, BAS BIOFLAME and SCAR EBA Programs and was supported by the Italian PNRA (Programma Nazionale di Ricerche in Antartide). We thank Enza D'Angelo for contributing to the compilation of the database, and Andrew Clarke and Lloyd Peck for comments on an earlier draft. We are grateful to the subject editor Ulrich Brose, Ethan White and an anonymous reviewer for constructive criticism that substantially improved the manuscript.

## REFERENCES

Addo-Bediako A, et al. 2002. Metabolic cold adaptation in insects: a large-scale perspective. – Funct. Ecol. 16: 332-338.

Blackburn, T.M. and Gaston, K.J. (eds) 2003. Macroecology: Concepts and Consequences.

British Ecological Society Annual Symposium Series. Blackwell Science.

Block, W. 1976. Oxygen uptake by Nanochestes antarcticus (Acari). – Oikos 27: 320-323.

Block, W. 1977. Oxygen consumption of the terrestrial mite *Alaskozetes antarcticus* (Acari: Cryptostigmata). – J. Exp. Biol. 68: 69-87.

Block, W. 1979. Oxygen consumption of the Antarctic springtail *Parisotoma octooculata* (Willem) (Isotomidae). – Rev. Ecol. Biol. Sol 16: 227-233.

Block, W. 1990. Cold tolerance of insects and other arthropods. – Philos. Trans. Royal Soc. B Biol. Sci. 326: 613-633.

Block, W. 1992. An annotated bibliography of Antarctic invertebrates (terrestrial and freshwater). – British Antarctic Survey.

Block, W. and Tilbrook, P.J. 1975. Respiration sudies on the Antarctic collembolan *Cryptopygus Antarcticus*. – Oikos 26: 15-25.

Block, W. and Tilbrook, P.J. 1978. Oxygen uptake by *Cryptopygus Antarcticus* (collembola) at South Georgia. – Oikos 30: 61-67.

Block, W. and Young, S.R. 1978. Metabolic adaptations of Antarctic terrestrial microarthropods. – Comp. Biochem. Physiol. 61A: 363-368.

Brown, J.H. et al. 1997. Allometric scaling laws in biology. – Science 278: 373.

Brown, J.H. et al. 2004. Toward a metabolic theory of ecology. – Ecology 85: 1771-1789.

Brown JH et al. 2005. Yes, West, Brown and Enquist's model of allometric scaling is both mathematically correct and biologically relevant. – Funct. Ecol. 19: 735-738.

Burnham, K.P and Anderson, D.R. 2002. Model selection and multimodel inference. A practical information theoretic approach. Second Edition. – Springer.

Cannon, R.J.C. and Block, W. 1988. Cold tolerance of microarthropods. – Biol. Rev. 63:23-77.

Chown, S.L. and Convey, P. 2006. Biogeography. Trends in Antarctic terrestrial and limnetic ecosystems. Antarctica as a global indicator. – In: Bergstom D.M., Convey, P., Huiskes H.L. (eds), Trends in Antarctic terrestrial and limnetic ecosystems. Antarctica as a global indicator. Springer, pp. 55-70.

Chown, S.L. and Gaston, K.J. 1999. Exploring links between physiology and ecology at macro-scales: the role of respiratory metabolism in insects. – Biol. Rev. 74: 87–120.

Chown, S.L. et al. 2007. Scaling of insect metabolic rate is inconsistent with the nutrient supply network model. – Funct. Ecol. 21: 282-290.

Convey, P. 1996. The influence of environmental characteristics on life history attributes of Antarctic terrestrial biota. – Biol. Rev. 71: 191–225.

Convey, P. 1997. How are the life history strategies of Antarctic terrestrial invertebrates influenced by extreme environmental conditions? – J. Therm. Biol. 22: 429-440.

Convey, P. 2006. Antarctic climate change and its influences on terrestrial ecosystems. – In: Bergstom D.M., Convey, P., Huiskes, H.L. (eds), Trends in Antarctic terrestrial and limnetic ecosystems. Antarctica as a global indicator. Springer, pp. 253-272.

Convey, P. et al. (2003) Soil arthropods as indicators of water stress in Antarctic terrestrial habitats? – Glob Change Biol 9: 1718-1730.

Duncan, R.P. et al. 2008. Testing the metabolic theory of ecology: allometric scaling exponents in mammals. – Ecology 88: 324–333

Garlaschelli, D. et al. 2003. Universal scaling relations in food webs. – Nature 423: 165-168.

Gillooly, J.F. et al. 2001. Effects of size and temperature on metabolic rate. – Science 293: 2248–2251.

Gillooly, J.F. et al. 2002. Effects of size and temperature on developmental time. – Nature 417: 70-73.

Glazier, D.S. 2005. Beyond the "3/4-power law": variation in the intra- and interspecific scaling of metabolic rate in animals. – Biol. Rev. Camb. Philos. Soc. 80: 611-662.

Goddard, D.G. 1977a. The Signy Island terrestrial reference sites: VIII. Oxygen uptake of some Antarctic prostigmatid mites (Acari Prostigmata). – Bull. Br. Antarct. Sur. 45: 101-115.

Goddard, D.G. 1977b. The Signy Island terrestrial reference sites: VI. Oxygen uptake of Gamasellus racovitzai (Trouessart) (Acari Mesostigmata). – Bull. Br. Antarct. Sur. 45: 1-11.

Hennion, F. et al. 2006. Physiological traits of organisms in a changing environment. – In: Bergstom D.M., Convey, P., Huiskes, H.L. (eds), Trends in Antarctic terrestrial and limnetic ecosystems. Antarctica as a global indicator. Springer, pp 129-160.

Hodkinson, I.D. 2003. Metabolic cold adaptation in arthropods: a smaller-scale perspective. – Funct. Ecol. 17: 562-567.

Huxley, J. S. 1932. Problems of relative growth. – Methuen.

Johnson, J.B. and Omland, K.S. 2004. Model selection in ecology and evolution. – Trends Ecol. Evol. 19: 101-108.

Kerkhoff A.J., Enquist B.J., 2008. Multiplicative by nature: Why logarithmic transformation is necessary in allometry. – J. Theor. Biol. 257: 519-521.

Kleiber, M. 1932. Body size and metabolism. – Hilgardia 6: 315–332.

Kozlowski, J. and Konarzewski, M. 2004. Is West, Brown andEnquist's model of allometric scaling mathematically correctand biologically relevant? – Funct. Ecol. 18: 283–289.

Makarieva, A.M. et al. 2005. Biochemical universality of living matter and its metabolic implications. – Funct. Ecol. 19: 547-557.

Makarieva, A.M. et al. 2008. Mean mass-specific metabolic rates are strikingly similar across life's major domains: Evidence for life's metabolic optimum. – PNAS 105: 16994-16999

Marquet, P.A. et al. 2004. Metabolic Ecology : linking individuals to ecosystems. – Ecology 85: 1794-1796.

McGaughran, A. et al., 2009. Temporal metabolic rate variation in a continental Antarctic springtail. – J. Insect Physiol. 55: 129-134.

Meehan, T.D. 2006. Mass and temperature dependency of metabolic rate in bitter and soil invertebrates. – Physiol. Biochem. Zool. 79: 878-884.

Osborne, J.W. and Overbay, A. 2004. The power of outliers (and why researchers should ALWAYS check for them). – Pract. Assess. Res. Eval. 9:6.

Packard, G.C. and Birchard, G.F. 2008. Traditional allometric analysis fails to provide a valid predictive model for mammalian metabolic rates. – J. Exp. Biol. 211: 3581-3587.

Packard, G. C., 2009. On the use of logarithmic transformations in allometric analyses. – J. Theor. Biol. 257: 515–518

Peck, LS. et al. 2006. Environmental constraints on life histories in Antarcic ecosystems: tempos timings and predictability. – Biol Rev 81: 75-109.

Peters, R.H. 1983. The Ecological Implications of Body Size. Cambridge University Press.

Precht, H. et al. 1973. Temperature and Life. – Springer.

Price, C.A. et al. 2009. Evaluating scaling models in biology using hierarchical Bayesian approach. Ecol. Lett. 12: 641-651.

Procter, D.L.C. 1977. Invertebrate respiration on Truelove Lowland. Truelove Lowland, Devon Island, Canada. – In: Bliss, L.C. (ed) A High Arctic Ecosystem. University of Alberta Press, pp. 383-393.

Prus, T. 1975. Bioenergetic budgets and balances. – In: Grodzinski, W., Klekowski, R.Z., Duncan, A. (eds), Methods for Ecological Bioenergetics, Blackwell Scientific, pp. 263–274.

R Development Core Team, 2006. R: A language and environment for statistical computing. R Foundation for Statistical Computing, Vienna, Austria. URL http://www.R-project.org

Ritz, C. and Streibig J.C. 2008. Nonlinear regression with R. – Springer.

Savage, V. M. et al. 2004. The predominance of quarter-power scaling in biology. – Funct. Ecol. 18: 257–282.

Sterner, R.W. and Elser J.J. 2002. Ecological Stoichiometry: The Biology of Elements from Molecules to the Biosphere. – Princeton University Press.

Turner, J. et al. 2005. Antarctic climate change during the last 50 years. – Int. J. Climatol. 25: 279-294.

Young, S.R. 1979. Effect of temperature change on the metabolic rate of an Antarctic mite. J. Comp. – Physiol. 131: 341-346.

Wall, D.H. 2007. Global Change tipping points: Above- and below-ground biotic interactions in a low diversity ecosystem. – Philos. Trans. Royal Soc. B Biol. Sci. 362: 2291-306.

Warton, D.I. et al. 2006. Bivariate line-fitting methods for allometry. – Biol. Rev. 81: 259–291.

West, G. B. et al. 1997. A general model for the origin of allometric scaling laws in biology. –Science 276: 122-126.

West, G. B. et al. 1999. The fourth dimension of life: fractal geometry and allometric scaling of organisms. – Science 284:1677–1679.

White, C.R. and Seymour, R.S. 2003. Mammalian basal metabolic rate is proportional to body mass 2/3. – PNAS 100: 4046–4049.

Table 1. Competing non-linear models fitted to non-transformed original data and ranked according to AIC criterion corrected for sample size. $I_0$, the scaling exponent $b$ and activation energy $E$ may vary among the four higher taxonomic groups [G] to which species belong (Collembola, Mesostigmata, Prostigmata, Cryptostigmata) or among the nine analysed species [S]. K is the number of estimated parameters (which includes estimation of residual $\sigma^2$), AICc is Akaike's Information Criterion corrected for sample size, Δ AIC is the difference in AIC between a model and the best-fitting model in the candidate set, which has a Δ AIC of 0; $R^2$ is the coefficient of determination.

| Model | $R^2$ | K | $AIC_c$ | Δ $AIC_c$ | Akaike's Weight |
|---|---|---|---|---|---|
| $I = I_0 M^{b[G]} e^{-E/kT}$ | 0.670 | 7 | -1854.70 | 0.00 | 0.303 |
| $I = I_0 M^{b} e^{-E[G]/kT}$ | 0.670 | 7 | -1854.67 | 0.03 | 0.298 |
| $I = I_0 \left[G\right] M^{b} e^{-E/kT}$ | 0.670 | 7 | -1854.62 | 0.07 | 0.292 |
| $I = I_0 M^{b[G]} e^{-E[G]/kT}$ | 0.682 | 10 | -1852.56 | 2.14 | 0.104 |
| $I = I_0 M^{b[S]} e^{-E/kT}$ | 0.672 | 12 | -1843.53 | 11.16 | 0.001 |
| $I = I_0 M^{b} e^{-E[S]/kT}$ | 0.672 | 12 | -1843.47 | 11.23 | 0.001 |
| $I = I_0 \left[S\right] M^{b} e^{-E/kT}$ | 0.671 | 12 | -1843.42 | 11.27 | 0.001 |
| $I = I_0 M^{b[S]} e^{-E[S]/kT}$ | 0.685 | 20 | -1827.90 | 26.80 | 0.000 |
| $I = I_0 M^{b} e^{-E/kT}$ | 0.532 | 4 | -1815.82 | 38.88 | 0.000 |
| $I = I_0 [G] M^{b}$ | 0.533 | 6 | -1811.68 | 43.01 | 0.000 |
| $I = I_0 [S] M^{b}$ | 0.537 | 11 | -1801.18 | 53.51 | 0.000 |
| $I = I_0 M^{b}$ | 0.537 | 3 | -1788.52 | 66.17 | 0.000 |

Table 2. Competing linear models fitted to log-transformed data and ranked according to AIC criterion corrected for sample size. $I_0$, the scaling exponent $b$ and activation energy $E$ may vary among the four higher taxonomical groups (G) to which species belong (Collembola, Mesostigmata, Prostigmata, Cryptostigmata) or among the nine analysed species (S). In allometric linear modelling, this variability can be accounted for by adding a linear term (S or G) and/or an interaction (:) term, namely S : log(M) or G : log(M) and S : (1/kT) or G : (1/kT). K, AICc, Δ AIC and $R^2$ are defined as in Table 1.

| Model | $R^2$ | K | $AIC_c$ | Δ $AIC_c$ | Akaike's Weight |
|---|---|---|---|---|---|
| $\log(I) = \log(I_0) + S + b\log(M) + (E / kT)$ | 0.827 | 12 | 281.94 | 0 | 0.996 |
| $\log(I) = \log(I_0) + S + S : b\log(M) + (E / kT)$ | 0.826 | 19 | 293.24 | 11 | 0.004 |
| $\log(I) = \log(I_0) + S + b\log(M) + S : (E / kT)$ | 0.819 | 19 | 297.88 | 16 | 0.000 |
| $\log(I) = \log(I_0) + G + b\log(M) + (E / kT)$ | 0.788 | 7 | 301.66 | 20 | 0.000 |
| $\log(I) = \log(I_0) + G + G : b\log(M) + (E / kT)$ | 0.788 | 10 | 305.45 | 24 | 0.000 |
| $\log(I) = \log(I_0) + S + b\log(M)$ | 0.790 | 11 | 305.76 | 24 | 0.000 |
| $\log(I) = \log(I_0) + G + b\log(M) + G : (E / kT)$ | 0.786 | 10 | 306.78 | 25 | 0.000 |
| $\log(I) = \log(I_0) + G + G : b\log(M) + G : (E / kT)$ | 0.786 | 13 | 310.88 | 29 | 0.000 |
| $\log(I) = \log(I_0) + S + S : b\log(M) + S : (E / kT)$ | 0.818 | 26 | 311.40 | 29 | 0.000 |
| $\log(I) = \log(I_0) + G + b\log(M)$ | 0.757 | 6 | 318.39 | 36 | 0.000 |
| $\log(I) = \log(I_0) + b\log(M) + (E / kT)$ | 0.736 | 4 | 326.84 | 45 | 0.000 |
| $\log(I) = \log(I_0) + b\log(M)$ | 0.697 | 3 | 343.83 | 62 | 0.000 |

Table 3. Estimated scaling exponent *b* (and 95 % confidence interval, CI) for relationships between individual metabolic rate and body mass. Non-linear and log-transformed models are compared. For log and non-linear models, G and [G] respectively indicates the effect on $I_0$ of the four taxonomical groups to which the nine analysed species (S) belong (Collembola, Mesostigmata, Prostigmata, Cryptostigmata). The species effect is indicated by S and [S]. These models assume that species or taxa do not differ in exponent or activation energy.

| Model | *b* | Lower 95 % CI | Upper 95 % CI |
|---|---|---|---|
| $I = I_0\left[G\right]M^b e^{-E/kT}$ | 0.721 | 0.533 | 0.909 |
| $\log(I) = \log(I_0) + G + b\log(M) + (E/kT)$ | 0.731 | 0.626 | 0.835 |
| $I = I_0\left[S\right]M^b e^{-E/kT}$ | 0.701 | 0.519 | 0.902 |
| $\log(I) = \log(I_0) + S + b\log(M) + (E/kT)$ | 0.696 | 0.591 | 0.801 |
| $I = I_0 M^b e^{-E/kT}$ | 0.546 | 0.420 | 0.671 |
| $\log(I) = \log(I_0) + b\log(M) + (E/kT)$ | 0.787 | 0.703 | 0.870 |
| $I = I_0[G]M^b$ | 0.621 | 0.413 | 0.829 |
| $\log(I) = \log(I_0) + b\log(M) + G$ | 0.732 | 0.620 | 0.844 |
| $I = I_0[S]M^b$ | 0.601 | 0.392 | 0.810 |
| $\log(I) = \log(I_0) + b\log(M) + S$ | 0.688 | 0.573 | 0.804 |
| $I = I_0 M^b$ | 0.496 | 0.355 | 0.636 |
| $\log(I) = \log(I_0) + b\log(M)$ | 0.787 | 0.698 | 0.877 |

Table 4. Parameter estimates for the best non linear model $I = I_0 M^{b[G]} e^{-E/kT}$, which assumes the allometric scaling exponent *b* may vary among the four major taxa to which the 9 analysed species belong

| Parameter | Estimate | Standard Error |
|---|---|---|
| $I_0$ = prefactor (normalization constant) | 12910 | 37280 |
| $E$ = activation energy | 0.488 | 0.071 |
| $b1$ = Cryptostigmata | 0.716 | 0.103 |
| $b2$ = Collembola | 0.634 | 0.081 |
| $b3$ = Mesostigmata | 0.803 | 0.091 |
| $b4$ = Prostigmata | 0.579 | 0.198 |

Figure Captions

Fig.1 Relationships between the temperature corrected metabolic rates *I,* measured in J $h^{-1}$, and body mass M measured in μg. According to the metabolic scaling theory (MST), the temperature correction is based on the factor exp(*E*/*kT*), where *E*, *k* and *T* are the activation energy, the Boltzmann's constant (8.62 x $10^{-5}$ eV $K^{-1}$), and absolute temperature respectively. The value of *E* was estimated from the best non linear model $I = I_0 M^{b[G]} e^{-E/kT}$. Multiple data are plotted for each species and each datum is marked by different symbol for the higher micro-arthropod taxon (Collembola, Mesostimagata, Prostigmata, Cryptostigmata) to which species belong. The wide scatter of points within taxa indicates a highly variable *I-M* scaling exponent. The thin and bold line show the MST (*b* = 3/4) and Euclidean (*b* = 2/3) predictions respectively, assuming that the allometric exponent (slope *b*), intercept (prefactor or so called normalization constant $I_0$) and activation energy (*E*) do not vary between and within species. Note that data were not log-transformed, even though axes are on a log scale.

Fig. 1

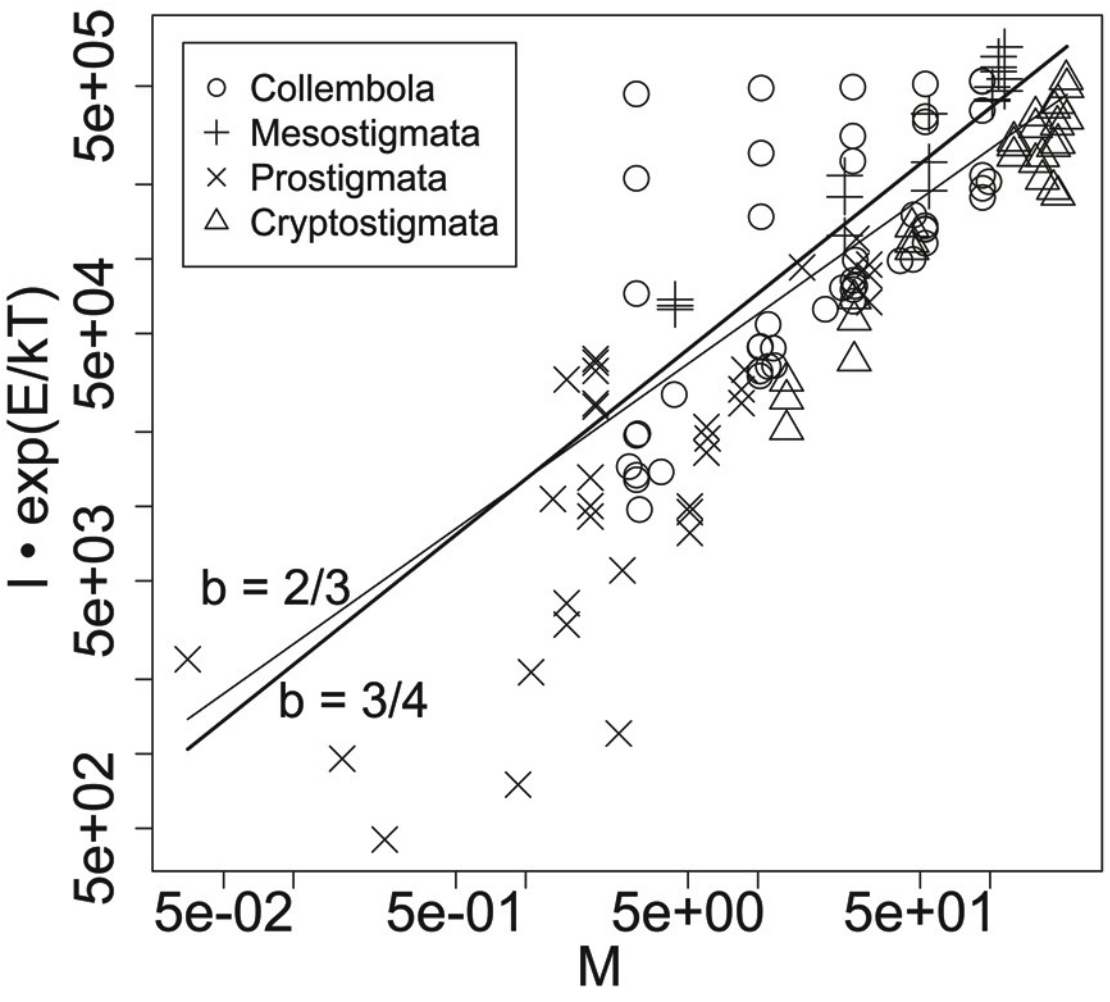

**Appendix 1.** Metabolic rate data for nine Antarctic microarthropod species

| Group | Species | Ref | Life Stage | Body mass (μg) | T (C°) | Metabolic rate (J $h^{-1}$) |
|---|---|---|---|---|---|---|
| Cryptostigmata | *Alaskozetes antarcticus* | 1 | I | 13.29 | 0 | 3.24E-05 |
| “ ” | “ ” | 1 | I | 13.29 | 5 | 3.98E-05 |
| “ ” | “ ” | 1 | I | 13.29 | 10 | 4.29E-05 |
| “ ” | “ ” | 1 | II | 25.99 | 0 | 3.91E-05 |
| “ ” | “ ” | 1 | II | 25.99 | 5 | 8.30E-05 |
| “ ” | “ ” | 1 | II | 25.99 | 10 | 1.45E-04 |
| “ ” | “ ” | 1 | III | 46.08 | 0 | 1.18E-04 |
| “ ” | “ ” | 1 | III | 46.08 | 5 | 1.60E-04 |
| “ ” | “ ” | 1 | III | 46.08 | 10 | 2.83E-04 |
| “ ” | “ ” | 1 | III | 126.65 | 0 | 2.63E-04 |
| “ ” | “ ” | 1 | III | 126.65 | 5 | 4.39E-04 |
| “ ” | “ ” | 1 | III | 126.65 | 10 | 6.02E-04 |
| “ ” | “ ” | 1 | IV | 156.97 | 0 | 2.48E-04 |
|  |  | 1 | IV | 156.97 | 5 | 5.55E-04 |
| “ ” | “ ” | 1 | IV | 156.97 | 10 | 7.13E-04 |
| “ ” | “ ” | 1 | IV | 187.67 | 0 | 1.93E-04 |
| “ ” | “ ” | 1 | IV | 187.67 | 5 | 4.17E-04 |
| “ ” | “ ” | 1 | IV | 187.67 | 10 | 8.54E-04 |
| “ ” | “ ” | 1 | IV | 196.21 | 0 | 1.83E-04 |
| “ ” | “ ” | 1 | IV | 196.21 | 5 | 4.3E-04 |
| “ ” | “ ” | 1 | IV | 196.21 | 10 | 8.93E-04 |
| “ ” | “ ” | 1 | IV | 168.02 | 0 | 2.12E-04 |
| “ ” | “ ” | 1 | IV | 168.02 | 5 | 3.80E-04 |
| “ ” | “ ” | 2 | IV | 214.18 | 5 | 7.22E-04 |
| “ ” | “ ” | 2 | IV | 212.72 | 5 | 7.78E-04 |
| “ ” | “ ” | 2 | IV | 213.71 | 10 | 7.72E-04 |
| “ ” | “ ” | 2 | IV | 194.85 | 10 | 7.30E-04 |
| Collembola | *Cryptopygus antarcticus* | 3 | I | 3.04 | 5 | 2.88E-05 |
| “ ” | “ ” | 3 | II | 10.26 | 5 | 6.49E-05 |
| “ ” | “ ” | 3 | III | 25.72 | 5 | 1.20E-04 |
| “ ” | “ ” | 3 | IV | 52.57 | 5 | 1.94E-04 |
| “ ” | “ ” | 3 | V | 98.81 | 5 | 3.01E-04 |
| “ ” | “ ” | 4 | I | 3 | 2 | 8.51E-05 |
| “ ” | “ ” | 4 | II | 10.323 | 2 | 1.74E-04 |
| “ ” | “ ” | 4 | III | 25.641 | 2 | 2.92E-04 |
| “ ” | “ ” | 4 | IV | 52.614 | 2 | 4.42E-04 |
| “ ” | “ ” | 4 | V | 92.574 | 2 | 6.13E-04 |
| “ ” | “ ” | 4 | I | 2.997 | 6 | 3.34E-04 |
| “ ” | “ ” | 4 | II | 10.323 | 6 | 4.22E-04 |
| “ ” | “ ” | 4 | III | 25.641 | 6 | 4.95E-04 |
| “ ” | “ ” | 4 | IV | 52.614 | 6 | 5.65E-04 |
| “ ” | “ ” | 4 | V | 92.574 | 6 | 6.28E-04 |
| “ ” | “ ” | 4 | I | 2.997 | 10 | 9.75E-04 |
| “ ” | “ ” | 4 | II | 10.323 | 10 | 1.03E-03 |
| “ ” | “ ” | 4 | III | 25.641 | 10 | 1.04E-03 |
| “ ” | “ ” | 4 | IV | 52.614 | 10 | 1.07E-03 |
| “ ” | “ ” | 4 | V | 92.574 | 10 | 1.10E-03 |
| “ ” | “ ” | 5 | I | 3 | 0 | 1.35E-05 |
| “ ” | “ ” | 5 | I | 3 | 5 | 2.84E-05 |
| “ ” | “ ” | 5 | I | 3 | 10 | 2.69E-05 |
| “ ” | “ ” | 5 | II | 10.2 | 0 | 3.40E-05 |
| “ ” | “ ” | 5 | II | 10.2 | 5 | 6.45E-05 |
| “ ” | “ ” | 5 | II | 10.2 | 10 | 7.39E-05 |
| “ ” | “ ” | 5 | III | 25.7 | 0 | 6.82E-05 |
| “ ” | “ ” | 5 | III | 25.7 | 5 | 1.20E-04 |
| “ ” | “ ” | 5 | III | 25.7 | 10 | 1.58E-04 |
| “ ” | “ ” | 5 | IV | 52.5 | 0 | 1.17E-04 |
| “ ” | “ ” | 5 | IV | 52.5 | 5 | 1.93E-04 |
| “ ” | “ ” | 5 | IV | 52.5 | 10 | 2.86E-04 |
| “ ” | “ ” | 5 | IV | 92.8 | 0 | 1.79E-04 |
| “ ” | “ ” | 5 | IV | 92.8 | 5 | 2.83E-04 |
| “ ” | “ ” | 5 | IV | 92.8 | 10 | 4.57E-04 |
| “ ” | “ ” | 6 | I | 3.83 | 5 | 2.02E-05 |
| “ ” | “ ” | 6 | II | 11.09 | 5 | 5.39E-05 |
| “ ” | “ ” | 6 | III | 23.12 | 5 | 1.12E-04 |
| “ ” | “ ” | 6 | I | 2.79 | 10 | 3.04E-05 |
| “ ” | “ ” | 6 | II | 11.78 | 10 | 7.78E-05 |
| “ ” | “ ” | 6 | III | 26.13 | 10 | 1.63E-04 |
| “ ” | “ ” | 6 | IV | 46.58 | 10 | 2.09E-04 |
| “ ” | “ ” | 6 | I | 3.09 | 20 | 4.03E-05 |
| “ ” | “ ” | 6 | II | 11.65 | 20 | 1.81E-04 |
| “ ” | “ ” | 6 | III | 19.51 | 20 | 2.60E-04 |
| “ ” | “ ” | 6 | IV | 40.98 | 20 | 4.08E-04 |
| Prostigmata | *Ereynetes macquariensis* | 7 | III | 1.5 | 0 | 3.29E-05 |
| “ ” | “ ” | 7 | IV | 2 | 0 | 3.85E-05 |
| “ ” | “ ” | 7 | IV | 2 | 5 | 5.17E-05 |
| “ ” | “ ” | 7 | IV | 2 | 10 | 5.46E-05 |
| Prostigmata | *Eupodes minutus* | 7 | IV | 2 | 0 | 3.98E-05 |
| “ ” | “ ” | 7 | IV | 2 | 5 | 3.71E-05 |
| Mesostigmata | *Gamasellus racovitzai* | 8 | I | 4.4 | 0 | 6.54E-05 |
| “ ” | “ ” | 8 | I | 4.4 | 5 | 9.17E-05 |
| “ ” | “ ” | 8 | I | 4.4 | 10 | 1.44E-04 |
| “ ” | “ ” | 8 | II | 23.65 | 0 | 1.80E-04 |
| “ ” | “ ” | 8 | II | 23.65 | 5 | 1.82E-04 |
| “ ” | “ ” | 8 | II | 23.65 | 10 | 4.56E-04 |
| “ ” | “ ” | 8 | III | 54.64 | 0 | 3.90E-04 |
| “ ” | “ ” | 8 | III | 54.64 | 5 | 3.61E-04 |
| “ ” | “ ” | 8 | III | 54.64 | 10 | 3.96E-04 |
| “ ” | “ ” | 8 | IV | 102.2 | 0 | 5.00E-04 |
| “ ” | “ ” | 8 | IV | 102.2 | 5 | 6.38E-04 |
| “ ” | “ ” | 8 | IV | 102.2 | 10 | 9.27E-04 |
| “ ” | “ ” | 8 | IV | 108.8 | 0 | 6.02E-04 |
| “ ” | “ ” | 8 | IV | 108.8 | 5 | 8.39E-04 |
| “ ” | “ ” | 8 | IV | 108.8 | 10 | 1.38E-03 |
| “ ” | “ ” | 8 | IV | 115.5 | 0 | 4.82E-04 |
| “ ” | “ ” | 8 | IV | 115.5 | 5 | 7.83E-04 |
| “ ” | “ ” | 8 | IV | 115.5 | 10 | 1.51E-03 |
| Prostigmata | *Nanorchestes antarcticus* | 9 | II | 2.61 | 5 | 8.08E-06 |
| “ ” | “ ” | 9 | III | 5.09 | 0 | 1.01E-05 |
| “ ” | “ ” | 9 | III | 5.09 | 5 | 1.39E-05 |
| “ ” | “ ” | 9 | III | 5.09 | 10 | 1.64E-05 |
| “ ” | “ ” | 9 | IV | 8.5 | 5 | 3.83E-05 |
| “ ” | “ ” | 9 | IV | 8.5 | 10 | 6.24E-05 |
| “ ” | “ ” | 9 | I | 0.035 | 5 | 3.53E-06 |
| “ ” | “ ” | 9 | II | 0.162 | 5 | 1.40E-06 |
| “ ” | “ ” | 9 | II | 1.055 | 5 | 3.13E-06 |
| “ ” | “ ” | 9 | II | 2.515 | 5 | 1.77E-06 |
| “ ” | “ ” | 9 | III | 8.748 | 5 | 5.23E-05 |
| “ ” | “ ” | 9 | III | 0.927 | 5 | 1.10E-06 |
| “ ” | “ ” | 9 | III | 6.07 | 5 | 2.76E-05 |
| “ ” | “ ” | 9 | III | 0.247 | 5 | 6.60E-07 |
| “ ” | “ ” | 9 | III | 1.31 | 5 | 1.57E-05 |
| Collembola | *Parisotoma octooculata* | 3 | I | 4.36 | 5 | 4.15E-05 |
| “ ” | “ ” | 3 | II | 11.07 | 5 | 7.97E-05 |
| “ ” | “ ” | 3 | III | 26.27 | 5 | 1.45E-04 |
| “ ” | “ ” | 3 | IV | 46.94 | 5 | 2.21E-04 |
| Prostigmata | *Stereotydeus villosus* | 7 | II | 6 | 5 | 3.07E-05 |
| “ ” | “ ” | 7 | II | 6 | 10 | 3.46E-05 |
| “ ” | “ ” | 7 | III | 15.5 | 10 | 1.94E-04 |
| “ ” | “ ” | 7 | IV | 26.6 | 0 | 7.71E-05 |
| “ ” | “ ” | 7 | IV | 26.6 | 5 | 1.77E-04 |
| “ ” | “ ” | 7 | IV | 26.6 | 10 | 1.97E-04 |
| “ ” | “ ” | 7 | IV | 30.18 | 0 | 6.76E-05 |
| “ ” | “ ” | 7 | IV | 30.18 | 5 | 1.39E-04 |
| “ ” | “ ” | 7 | IV | 30.18 | 10 | 1.80E-04 |
| Prostigmata | *Tydeus tilbrooki* | 7 | III | 1.5 | 0 | 3.36E-06 |
| “ ” | “ ” | 7 | III | 1.5 | 10 | 8.52E-06 |
| “ ” | “ ” | 7 | IV | 1.9 | 0 | 1.32E-05 |
| “ ” | “ ” | 7 | IV | 1.9 | 5 | 1.47E-05 |
| “ ” | “ ” | 7 | IV | 1.9 | 10 | 1.91E-05 |

Ref., reference: 1, Block 1977; 2, Young 1979; 3, Block 1979; 4, Procter, 1977; 5, Block and Tilbrook, 1975; 6, Block and Tilbrook 1978; 7, Goddard 1977a; 8, Goddard 1977b; 9, Block, 1976; Life stage: I, very young; II, young; III, sub-adult; IV, adult 1; V adult 2.

**References**

1, Block, W. 1977..Oxygen consumption of the terrestrial mite *Alaskozetes antarcticus* (Acari: Cryptostigmata). – J. Exp. Biol. 68: 69-87.

2, Young, S.R. 1979. Effect of temperature change on the metabolic rate of an Antarctic mite. – J. Comp. Physiol. 131: 341-346.

3, Block, W. 1979. Oxygen consumption of the Antarctic Springtail *Parisotoma octooculata* (Willem) (Isotomidae). – Rev. Ecol. Biol. Sol 16: 227-233.

4, Procter, D.L.C. 1977. Invertebrate respiration on Truelove Lowland. Truelove Lowland, Devon Island, Canada. – In: Bliss LC (ed) A High Arctic Ecosystem. University of Alberta Press, pp 383-393.

5, Block, W. and Tilbrook, P.J. 1975. Respiration sudies on the Antarctic collembolan *Cryptopygus Antarcticus*. – Oikos 26: 15-25.

6, Block, W. and Tilbrook, P.J. 1978. Oxygen uptake by *Cryptopygus Antarcticus* (collembola) at South Georgia. – Oikos 30: 61-67.

7, Goddard, D.G. 1977a. The Signy Island terrestrial reference sites: VIII. Oxygen uptake of some Antarctic prostigmatid mites (Acari Prostigmata). – Bull Br Antarc Surv 45: 101-115.

8, Goddard, D.G. 1977b. The Signy Island terrestrial reference sites: VI. Oxygen uptake of *Gamasellus racovitzai* (Trouessart) (Acari Mesostigmata). – Bull Br Antarct Surv 45: 1-11.

9, Block, W. 1976. Oxygen uptake by *Nanochestes antarcticus* (Acari). – Oikos 27: 320-323.